\documentstyle[12pt]{article}
\begin{document}
\begin{flushright}
UT-Komaba 97-1\\
\end{flushright}
\vspace{1cm}
\begin{center} 
{\Large{\bf  Fluctuation Effects of Gauge Fields \\
in the Slave-Boson t-J Model
}}
\vfill
 {\Large  Ikuo Ichinose$^{\star}$,\footnote{e-mail address: 
ikuo@hep1.c.u-tokyo.ac.jp}
  Tetsuo Matsui$^{\dagger}$,\footnote{e-mail address: 
matsui@phys.kindai.ac.jp}
   and Kazuhiko Sakakibara$^{\ast}$\footnote{e-mail address: 
sakaki@center.nara-k.ac.jp}}\\
\vskip 0.2in
 $^{\star}$Institute of Physics, University of Tokyo, Komaba, Tokyo, 153 Japan  \\
 $^{\dagger}$Department of Physics, Kinki University, Higashi-Osaka,  577 Japan \\
$^{\ast}$Department of Physics, Nara National College of Technology, Yamatokohriyama,
639-11 Japan
\vfill

\end{center}

\begin{center} 
\begin{bf}
Abstract
\end{bf}
\end{center}

We present a quantitative study of the charge-spin separation (CSS) phenomenon 
in a  U(1)  gauge theory of the t-J model of high-T$_C$ superconductors.
We calculate the critical temperature $T_{\rm CSS}$ of a 
confinement-deconfinement phase transition as a function of the hole doping 
$\delta$, below which the CSS takes place. The fluctuations
of gauge field are so large that $T_{\rm CSS}$ is reduced to
about 10 \%  of  its mean-field value.

\vfill
\eject

The anomalous properties of the metallic phase of cuprate 
superconductors,  like the linear temperature ($T$) 
dependence of resistivity, have called for  theoretical explanations.
The idea of charge-spin separation (CSS) of electrons into holons and spinons,
introduced by Anderson\cite{css},
seems to be an interesting possibility for  them.

Straightforward mean-field (MF) theories of the t-J model
in the slave-boson (SB) or slave-fermion representation are compatible 
with the CSS since holons and spinons  acquire their own 
hopping amplitudes through nonvanishing MF's.
Fluctuations around such MF's are described by a U(1) gauge theory.
Effects of these  fluctuations have been studied
within perturbative approaches. 
In Ref.\cite{ln} the resistivity is 
calculated to show a linear-$T$ dependence.
From a more general point of view, a system of nonrelativistic fermions 
coupled with a U(1) gauge field is studied by perturbative 
renormalization-group (RG), showing the non-Fermi-liquid-like 
behavior\cite{nw}. 
All these analyses assume that gauge-field fluctuations are ``{\em small} "
and can be handled by perturbative or RG methods.

In Ref.\cite{im1}, we developed a general formalism to
study such gauge fields in a {\em non-perturbative} manner, in order to
see whether and when these perturbative analyses are validated.
In Ref.\cite{im2}, we applied that formalism to the t-J model.
Our method  was  motivated by the success of the  Ginzburg-Landau (GL) theory 
of  the microscopic Bardeen-Cooper-Schriefer's (BCS) model, which, 
taking a form of the XY spin model,   
is useful for studying superconducting phase transition, 
dynamics of vortices, etc.


In contrast to the XY model for the BCS model, 
the resulting GL theory
for the t-J model takes  a form of lattice gauge theory,
reflecting the local gauge symmetry  in the slave-particle
representation, 
so the  gauge-theoretical approach is quite useful\cite{gauge}. 
  Gauge dynamics generally exhibits two phases; confinement phase with
large fluctuations of gauge fields and deconfinement phase with small fluctuations.
In the language of gauge theory, we characterized CSS as  a deconfinement 
phenomenon of holons and spinons\cite{im1}.
 By using the knowledge of lattice gauge theory,
we showed that there exists a confinement-deconfinement (CD) phase transition
at some critical temperature $T_{\rm CSS}$. At $T < T_{\rm CSS}$,
the deconfinement phase is realized, where holons and spinons are deconfined,
hence  the CSS takes place and the anomalous behavior is expected,  
while at $T > T_{\rm CSS}$,
the confinement phase is realized where holons and spinons are confined
in electrons and the usual Fermi-liquid picture of electrons should be applicable.

In this paper, we calculate $T_{\rm CSS}$ quantitatively 
following Ref.\cite{im2}. 
The result should be compared with the experimental indications of 
the onset $T$, below  which the anomalous behavior is observed.

In the SB representation of the t-J model, 
the electron operator $C_{x\sigma}$ [$x$; site, $\sigma (=1,2)$; spin] 
is written as $C_{x\sigma} = b^{\dagger}_x f_{x\sigma}$, where
 $f_{x\sigma}$ are the  fermionic spinon operators and $b_x$ are the bosonic 
holon operators. 
The physical states must satisfy the local constraint 
$(f^{\dagger}_{x 1}f_{x 1} + f^{\dagger}_{x 2}f_{x 2} + b^{\dagger}_x b_x-1)
|\mbox{phys}\rangle=0$ to avoid double occupancies of electrons.
The Hamiltonian on a  two-dimensional lattice is given by  
\begin{eqnarray}
H&=&-t\sum_{x,i,\sigma} (b^{\dagger}_{x+i}f^{\dagger}_{x \sigma}f_{x+i,\sigma}b_x+\mbox{H.c.})\nonumber\\
&-&\frac{J}{2} \sum_x(\sum_{\sigma}  f^{\dagger}_{x \sigma}\tilde{f}_{x+i,\sigma}) 
 (\sum_{\sigma'} \tilde{f}^{\dagger}_{x+i,\sigma'}f_{x \sigma'}) +H_{\mu}, 
\label{Hsb}
\end{eqnarray}
where $i (=1,2)$ is the direction  index,
$\tilde{f}_{x \sigma} \equiv 
\sum_{\sigma'}\epsilon_{\sigma \sigma'}f^{\dagger}_{x \sigma'}\;
 (\epsilon_{12}= -\epsilon_{21}=1 $ ), and
$H_{\mu} =  -\mu_f\sum f^{\dagger}_{x \sigma}f_{x \sigma}  
 -\mu_b\sum b^{\dagger}_xb_x. $
We relaxed the local constraint to the global one. The chemical potentials 
$\mu_{b}$ and $\mu_{f}$ are chosen as 
$\langle b^{\dagger}_x b_x\rangle = \delta$  
and  $  \langle\sum_{\sigma} f^{\dagger}_{x \sigma} f_{x \sigma}\rangle 
= 1 - \delta$. 

By using the Hubbard-Stratonovich transformation in the path-integral formalism,
we rewrite (\ref{Hsb}) in the  form of MF theory proposed in Ref.\cite{ul}.
Explicitly, by introducing auxiliary complex {\em link variables}, $\chi_{x i}$
and $D_{x i}$ which are {\it fluctuating} ``MF's", we get
\begin{eqnarray}
H&=&\sum_{x,i} \Big[ {3J\over 8}|\chi_{x i}|^2+{2\over 3J}|D_{x i}|^2
\Big] \nonumber\\
&-&\sum_{x,i}\Big[\chi_{x i}\Big({3J\over 8}\sum_{\sigma} f^{\dagger}_{x+i,\sigma}
f_{x \sigma} 
+t b^{\dagger}_{x+i}b_x\Big) +\mbox{H.c.}\Big]  \nonumber \\
&-&{1\over 2}\sum_{x,i,\sigma} 
[ D_{x i} f^{\dagger}_{x \sigma}\tilde{f}_{x+i,\sigma} +\mbox{H.c.} ]
+H_4 +H_{\mu}, 
\label{Hdec1}
\end{eqnarray}
where $H_4  =  {8t^2 \over 3J}\sum b^{\dagger}_{x+i}b_{x+i}b^{\dagger}_xb_x  
 - {3J\over 8}\sum f^{\dagger}_{x+i,\sigma}f_{x+i,\sigma'}
f^{\dagger}_{x \sigma'}f_{x \sigma}$. 
The partition function $Z(\beta) \; [\beta \equiv(k_B T)^{-1}]$  is given by
\begin{eqnarray}
Z &=& \int[db][df][d\chi][dD]\exp(A),\nonumber\\
A&=& \int_0^{\beta}d\tau [-\sum_x(\bar{b}_x\dot{b}_x+
\sum_{\sigma}\bar{f}_{x\sigma}\dot{f}_{x\sigma}) -H].
\label{z}
\end{eqnarray} 
From (\ref{z}),  one gets the relations, 
$\langle\bar{\chi}_{x i}\rangle = \langle\sum f^{\dagger}_{x+i,\sigma}
f_{x \sigma} +(8t/3J) b^{\dagger}_{x+i}b_x \rangle,$ 
$\langle\bar{D}_{x i}\rangle =(3J/4) \langle\sum f^{\dagger}_{x\sigma}\tilde{f}_{x+i,\sigma}\rangle$.
So $\chi_{x i} $ is the hopping amplitude
of holons and spinons, while  $D_{xi}$ is the amplitude of resonating 
valence bonds (RVB) of antiferromagnetism.

In Ref.\cite{im2}, we derived an equation that determines $T_{\rm CSS}$.
Here, we summarize the basic steps (I-IV) leading to that equation.

(I) We parametrize $\chi_{xi}$ and $D_{xi}$ a la ``uniform RVB state" as 
$\chi_{xi}=\chi U_{xi}, \;D_{xi}=D_{i}V_{xi}$, where $U_{xi}, V_{xi}\in U(1)$,
and  $\chi$, $D_{1}= -D_{2} =  D$  are link-independent 
constants, focusing on the U(1) phase dynamics. 
Eq.(\ref{Hdec1}) is invariant under a  time-independent U(1) gauge 
transformation, $(f_{x\sigma},b_x) \rightarrow 
e^{i\theta_x}(f_{x\sigma},b_x)$, while the above U(1) variables transform as 
$U_{xi}  \rightarrow e^{i\theta_x}U_{xi}e^{-i\theta_{x+i} }$ and  
$V_{xi}  \rightarrow e^{i\theta_x}V_{xi}e^{i\theta_{x+i} } $.
Therefore they are two different kinds of  lattice gauge variables. 

(II) We integrate over $f_{x\sigma}$ and $b_{x}$ by using 
the hopping expansion, an expansion 
in powers of $\chi_{xi}$ and $D_{xi}$.
By keeping the leading contributions (i.e., neglecting 
the plaquette terms, etc.),  we obtain the following effective 
 lattice gauge theory as the  GL theory of the t-J model;  
\begin{eqnarray}
Z & =& \int[dU][dV]\exp(A_{\rm eff}),\nonumber\\
A_{\rm eff}&=& \int_{0}^{\beta} d\tau \sum_{x,i}[-a_U 
|\dot{U}_{xi}|^2 - a_V |\dot{V}_{xi}|^2 ]),
\label{eq:zuv}
\end{eqnarray}
where $a_{U,V}$ are known functions of $\chi,D,\mu_{b,f}$ and $T$.

(III) Following the Polyakov-Susskind approach\cite{ps}
 to the CD transition we  map
(\ref{eq:zuv}) into an  {\it anisotropic}
2D classical XY spin model; 
\begin{eqnarray}
 Z&=&\int^{\pi}_{-\pi} \prod_{x} \frac{d \alpha_{x}}{2\pi} 
 \exp[\sum_{x,i}\{J_1\cos (\alpha_{x+i}-\alpha_x)
\nonumber\\
&+& J_2 \sum_{x,i} \cos 
(\alpha_{x+i} + \alpha_x)\}], 
\label{eq:zxy1}
\end{eqnarray}
where 
$  J_{1} \equiv   \beta^{-1} a_{U}  ,  J_{2} \equiv   \beta^{-1}  a_{V} $. 
It is obvious that $J_2 \propto D^2$ is an anisotropic parameter.
The spin-spin correlation $C(|x|) \equiv \langle\exp(i\alpha_x) \exp(-i\alpha_0)\rangle$
is expressed as $\exp(-\beta W(|x|))$, 
where $W(R)$ is the potential energy
of a pair of oppositely charged gauge sources separated by $R$. 

(IV) We investigated the spin model (\ref{eq:zxy1}) both by the 
effective action in large-spatial-dimensions \cite{im2} and the Monte Carlo simulation\cite{sm}.
From both analyses, we found that that there occurs an order-disorder phase 
transition along  the line $ J_1  + J_2 \simeq 1$, which is  
  identified  with the CD phase transition.
The condition  $J_1 + J_2 \simeq 1$ gives rise to the following  
 equation that determines $T_{\rm CSS}$\cite{cor};
\begin{eqnarray}
(k_B T_{\rm CSS})^2 \simeq \frac{\chi^2(T_{\rm CSS})} {4\pi^2}[ \frac{1}{2}
(\frac{3J}{8})^2 (1-\delta^2)+t^2\delta(1+\delta)] \nonumber\\
+ \frac{D^2(T_{\rm CSS})}{32F^3}[ 2\delta-(1-\delta)^2 F(F+1) ] ,
\label{tcd}
\end{eqnarray}
where $F \equiv \ln((1+\delta)/(1-\delta))$. 
There are three phases of the spin dynamics  (\ref{eq:zxy1}) 
and corresponding gauge 
dynamics\cite{mix}, which are summarized as follows.

$\bullet$  Disordered Phase  at $J_1 + J_2 < 1$:\\
\noindent $C(R) \sim \exp(-\gamma R)$ with some finite constant $\gamma$, and
so $W(R) \propto R$ is a confining potential.
The gauge system is in the {\em confinement  phase} ($\langle U_{xi}
\rangle = \langle V_{xi}\rangle =0$), where only 
the gauge-invariant neutral objects, like elecrons, appear.
 
$\bullet$ Ordered Phase   at  $  J_1+J_2  > 1, J_2 \neq 0$:\\
\noindent  
The $Z_2$ symmetry of (\ref{eq:zxy1}) is spontaneously broken.
Excitations from the ground state $\alpha_x=0$
(mod $\pi$) are {\em massive} spin waves with finite energy gap $m$;  
$C(R) \rightarrow const. + \exp(-m R)$.
Then, $W(x) \simeq \exp (-m|x|)$. This  is the {\em  deconfinement}
($\langle U_{xi}\rangle,\langle V_{xi}\rangle\neq 0$) {\em Higgs phase }
 where the gauge fields acquire  a mass $m$.

$\bullet $ Quasi-Ordered Phase at $J_1 > 1,J_2=0$:\\ 
\noindent
The model is O(2) symmetric and the Kosterlitz-Thouless  analysis
applies; $C(R) \simeq R^{-\eta}$, $W(R)$ $ \simeq \ln (R)$.
This is the {\em deconfinement Coulomb phase} with  massless gauge fields.
  
To determine $T_{\rm CSS}$, we first calculate  $\chi$, 
$D$, $\mu_{b,f}$ as functions of $T$ and $\delta$ by a straightforward MF theory 
that totally ignores 
gauge-field fluctuations \cite{mf}.
In Figs. 1 and 2, we summarize the results of such a MF theory with 
$J = 0.1 eV, t = 0.3 eV$.
Fig.1 shows  $J_1+J_2$
as a function of $T$ for various $\delta$'s.
In Fig.2, the curves $T_{\chi}$ and $T_{\rm RVB}$  show
the value of $T$ at which $\chi$ and $D$ vanish, respectively.
$T_{\rm BC}$ is the onset $T$, below which the Bose condensation $\langle b_x\rangle \neq 0$ takes place.  To calculate $T_{\rm BC}$ we included  
a weak but finite three-dimensionality, 
$t' = \alpha \; t, \alpha \simeq 10 ^{-2}$, while $\chi,D$ are calculated with
$\alpha = 0$. 
Using these $\chi $, $D $, we can solve (\ref{tcd}) to 
 obtain $T_{\rm CSS}$ as a function of $\delta$ as plotted in Fig.2.
We find that $T_{\rm CSS}$ is smaller by
factor $\sim 1/10$ than its mean-field value $T_{\chi}$. 
Fig.3 shows this relation together with $\chi(T)$.
Figs.2,3 exhibit  that the effects of gauge 
fluctuations on $T_{\rm CSS}$ are 
so strong that the MF estimation is quantitatively incorrect.
This result is welcome since the experimental signal of the onset of CSS
is around $T \sim$ several-hundred $K$.

To check the stability of $T_{\rm CSS}$, let us suppose that
the condition for the transition would be $J_1 + J_2 =A$ with
$A = 2$ instead of $A=1$,  due to, say, the higher-order terms in the hopping 
expansion. Fig.1 shows that $T_{\rm CSS}$ would shift only  $20 \sim 30\%$ 
from its present value since $J_1 + J_2$ decreases very rapidly
up to $J_1 + J_2 \simeq 0.5 $.  This reflects the fact that
$\chi(T)$ is almost saturated 
in this {\it low} $T$ region, as shown in Fig.3.

The two-body term $H_4$ in (\ref{Hdec1}),  neglected in the MF theory, is 
necessary to keep the equivalence to the original t-J model. 
In the leading-order of perturbation theory, it reduces $T_{\rm CSS}$
 furthermore.
Also one may ask how  $T_{\rm CSS}$ changes
in  other decoupling schemes \cite{shf}.
Although $T_{\rm CSS}$ may certainly shift,  
the importance of large fluctuations should remain true.   
We shall report on  these points in future publications.


The concept of CSS   
is very drastic, and it is difficult to handle it
by the conventional approaches in the condensed matter physics.
However,  the notion of gauge theory affords us to understand it
as a familiar phenomenon, like the CD phenomenon of quarks and gluons
in the color gauge theory of strong interactions. 
We note that the  gauge theory of separation  phenomena used  here
found yet another application; the particle-flux separation 
in the fractional quantum Hall effects\cite{pfs}.

\eject

\eject

FIGURE CAPTIONS.

\begin{flushleft}
Fig.1. $J_1 + J_2$ v.s. $T$.
The CSS sets in at $J_1 + J_2  \simeq 1$.\\

\vspace{0.5cm}

Fig.2.  MF results and $T_{\rm CSS}$.
Along  $T_{\chi}$ and $T_{\rm RVB}$, $\chi$   and
$D$ vanish, respectively. $T_{\rm BC}$ is the
onset $T$ of  Bose condensation, $\langle b_x\rangle \neq 0$.
$T_{\rm CSS}$ is the critical $T$ of the CD transition  
calculated by Eq.(\ref{tcd}), {\it below} which the CSS takes place. 
The dashed line starting from P is an {\it  expected} $T$
along which $\langle D_{xi} \rangle$ vanishes, i.e., $T_{\rm RVB}$ renormalized 
by  gauge fluctuations $\langle V_{xi}\rangle$. 
Similarly, the renormalized $T_{\chi}$ is just $T_{\rm CSS}$.
There are five phases:\\
(i) Strange Metal Phase:  Deconfinement-Coulomb;\\
(ii) Spin-Gap Phase:   Deconfinement-Higgs with $D \neq 0$;\\
(iii) Fermi-Liquid Phase:  Deconfinement-Higgs 
with $\langle b_x\rangle\neq 0$; \\ 
(iv) Superconducting Phase: Deconfinement-Higgs 
with $D,  \langle b_x\rangle  \neq 0$;\\
(v) Electron Phase: Confinement {\it above} $T_{\rm CSS}$. \\

\vspace{0.5cm}

Fig.3.  
$\chi(T)$ at $\delta = 0.15$. The overall structure
of $\chi(T)$ and the relation $T_{\rm CSS} \sim T_{\chi}/10$ remain similar
for other values of $0 < \delta < 1$.
\end{flushleft}


\begin{thebibliography}{99}
\bibitem{cor}
The corresponding Eq.(3.30) in Ref.\cite{im2} should be corrected as 
(\ref{tcd}). This reflects the errata; The L.H.S. of the first line in Eq.(3.8) should be multiplied by  a factor 2.  Also the minus signs of $\mu_{f,b}$ in the Hamiltonian (3.2,3) should be reversed. To obtain (\ref{tcd}), we used
$\mu_{b,f}(\delta)$ determined by the hopping expansion, not by the MFT.
\bibitem{mix} One might conceive the possibility of mixed  dynamics;
 the $U$ field (say, $b-\bar{b}$ channel) is in the  confinement phase
whereas the $V$ ($b-b$ channel) field is in the deconfinement
 (or otherway around).
However, a MF theory and strong-coupling expansion for (\ref{eq:zxy1}) 
shows $C(|x|)$ and $\langle\exp(i\alpha_x) \exp(i(-)^{x_1 +x_2}
\alpha_0)\rangle$
behave similarly, denying such a possibility.
\bibitem{mf}
Our MF theory neglected $H_4$ of (\ref{Hdec1}) as in  \cite{ul}.
 We assumed the uniform RVB, without setting the spatial phases
as variational parameters as in their case.
\end{thebibliography}
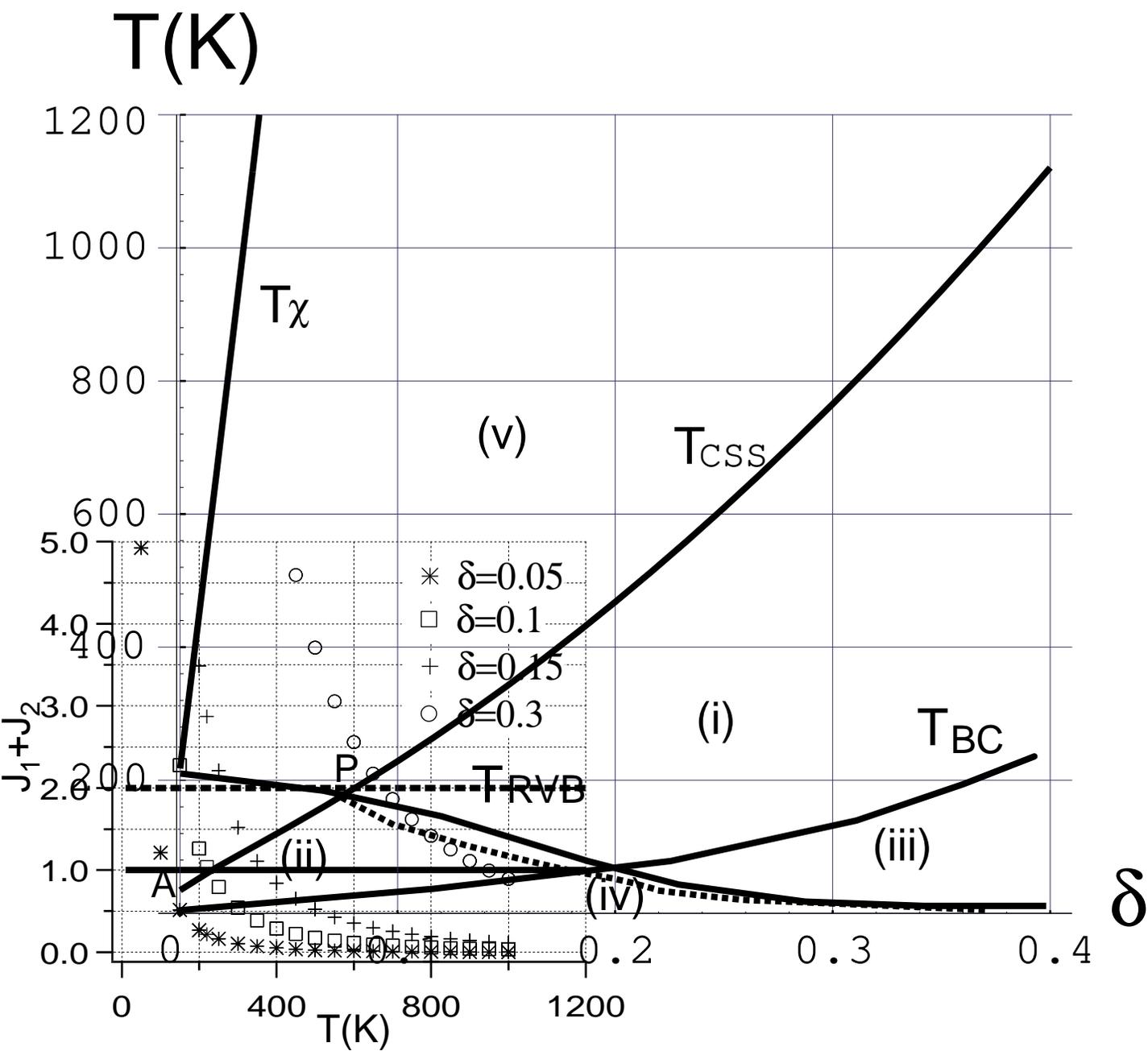
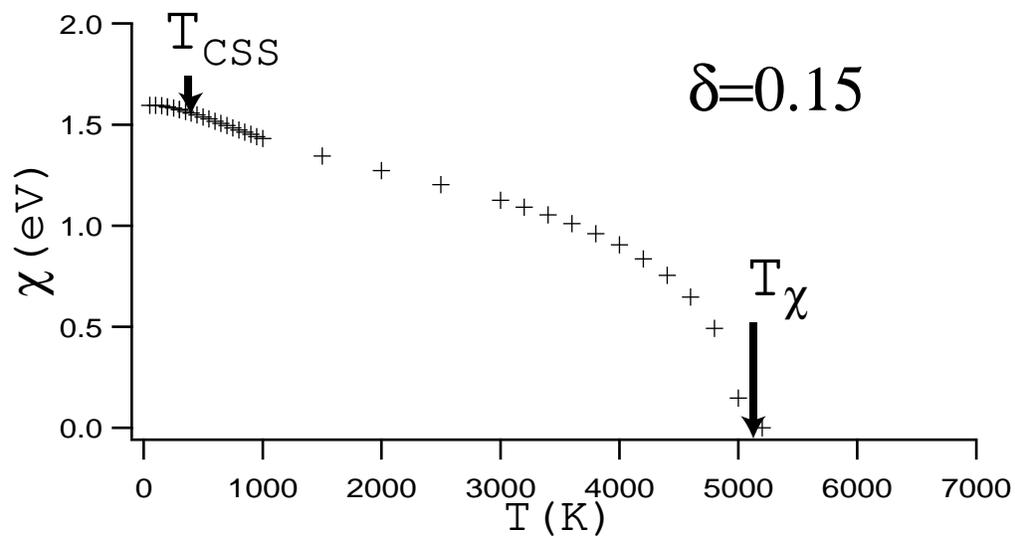
\end{document}